\documentstyle[aps,epsf]{revtex}

\begin{document}

\title{Instanton effective action formalism for 2-point and 4-point functions}

\author{F. S. Roux and T. Torma \\ {\it Department of Physics, University
of Toronto \\ Toronto M5S1A7, CANADA } }

\maketitle

\begin{abstract} 
We provide an instanton effective action formalism in terms of which one can
investigate the nonperturbative generation of 2-point and 4-point functions.
\end{abstract}


\section{Introduction}
\label{intro}

To understand the physics above the electroweak scale one needs to understand
nonperturbative dynamics.  If Nature turns out to be supersymmetric not far above the
electroweak scale then one seems to require nonperturbative dynamics to understand
how supersymmetry breaks.  Otherwise, a process of dynamical symmetry breaking is
required to avoid unnatural fine tuning.

An important feature of nonperturbative dynamics, is the formation of composite order
parameters -- fermion n-point functions that are generated nonperturbatively through
strong dynamics.  Much work has been done on 2-point functions because of the role
they play in chiral symmetry breaking, which is not only relevant to QCD, but also to
physics beyond the standard model.

Recently \cite{r_h,r_rth} it was pointed out that 4-point functions may play an
important role in flavor scale physics, high above the electroweak scale. The argument
is as follows.  The situation at such high scales is that fermions are (effectively)
massless so that a chiral symmetry exists. Above the electroweak scale there would not
be any 2-point functions that involve fermions with electroweak quantum numbers,
because these would break the electroweak symmetry. The same applies to any other
n-point function that breaks the electroweak symmetry.  There are however some
4-point functions involving fermions with electroweak quantum numbers that are allowed
at scales above the electroweak scale.  One can distinguish five such 4-point functions
on the basis of the chiralities of their external lines:\ three are chirality
preserving and two are chirality changing.  The chirality preserving 4-point functions
can obviously respect any chiral symmetry.  The chirality changing 4-point functions
can be invariant under a chiral isospin symmetry, which may contain the $SU(2)_L$ of
the electroweak symmetry, but these 4-point functions break larger chiral symmetries.
If it can be generated nonperturbatively, such a 4-point function would signal the
following breaking pattern
\begin{equation}
SU(N_f)_R \times SU(N_f)_L \rightarrow SU(2)_R \times SU(2)_L \times SU(N_f/2)_V .
\label{ie0}
\end{equation}
(The subscript $V$ indicates a vectorial symmetry.) Hence, these chirality changing
4-point functions act as order parameters for this partial breaking of chiral
symmetries, which may occur at scales high above the electroweak scale. (Note that all
fermions transform nontrivially under chiral isospin, which implies that masses are
not allowed.)

The significance of this is that the partial chiral symmetry breaking gives a new
scale high above the electroweak scale. The surviving chiral isospin symmetry
can contain the $SU(2)_L$ of the electroweak symmetry, which means that this mechanism
can exist for standard model fermions. It is therefore not necessary to introduce new
non-standard model fermions to produce flavor scale physics.  As a result dynamical
models may become simpler.

For 4-point functions to act as order parameters they must appear at higher scales
than 2-point functions do. One can imagine that for certain combinations of number of
colors, $N_c$, and number of flavors, $N_f$, the critical coupling for the formation of
2-point functions may be higher than the critical coupling for the formation of
4-point functions.  In an asymptotically free theory the 4-point functions would then
be formed at a higher scale than the 2-point functions. The separation between these
scale may be larger for {\it large numbers of flavors} than it would be in QCD-like
theories where the coupling runs relatively quickly.  Under these circumstances one
would have the breaking pattern of (\ref{ie0}) at the higher scale where the 4-point
functions form and lower down, when 2-point functions form, the chiral isospin
symmetry would break to a vectorial isospin symmetry. In some cases, for large enough
numbers of flavors, an infrared fixed point may prevent the coupling from reaching the
critical value for the formation of 2-point functions. All this is of course still
speculation. The question therefore is, under what circumstances ($N_c$ and $N_f$), if
any, would a nonperturbatively generated 4-point function appear at higher scales
than 2-point functions do?

One can start by assuming that the favorable situation exists for 4-point functions to
form at a much higher scale than 2-point functions. This means that one can assume
that fermions remain massless.  It allows one to formulate an effective action for
these 4-point functions, making the chiralities of the fermions explicit.  Thus one can
distinguish the above mentioned five different 4-point functions.  Such an analysis
was done in \cite{r_rth}, considering only gauge exchanges.  The outcome was that in
the limit of a large number of colors, none of the five 4-point functions can be
generated nonperturbatively.  Since the analysis in \cite{r_rth} excluded instanton
dynamics, the question naturally arises whether 4-point functions can perhaps be
generated through instanton dynamics.

The instanton \cite{r_bpst,r_t} is an important nonperturbative effect that appears
in non-Abelian gauge theories.  The relevence of instantons in the formation of
2-point functions and, by implication, in chiral symmetry breaking has been realized
long ago \cite{r_c,r_cc}.  Subsequently many authors elucidated various aspects of
this mechanism for chiral symmetry breaking.  For reviews see \cite{r_d,r_s}.

Unlike the 2-point function case, calculations of the nonperturbative generation of
4-point functions through instantons has received very little if any attention so
far.  At present there does not seem to exist a formalism in which one can treat this
problem.  Therefore the purpose of this paper is to provide an effective action
formalism for the calculation of the instanton contribution to the formation of
4-point functions.  Because such a formalism requires as part of its derivation a
consistent treatment of 2-point functions, it also provides the means to calculate the
instanton contribution to their formation.  Our formalism provides a consistent
framework for deriving the gap equations of the relevant n-point functions through the
stationarity of the effective action.  The instanton contributions to the effective
action are represented as diagrams that are generated from Feynman rules.  These rules
include a nonlocal instanton vertex for the fermion zeromodes and a fermion
propagator.  The nonlocal instanton vertex differs from the usual effective 't Hooft
vertex \cite{r_t} in the sense that, unlike the latter, it is valid at distance scales
smaller than the instanton size.

The coupling strengths in our diagrammatic language are given by:\ the sizes of the
n-point functions and the coefficient of the instanton vertex, which depends on the
gauge coupling, number of colors and number of flavors in a complicated way. For a
specific size of the coefficient of the instanton vertex, the sizes of the actual
n-point functions determine whether the expansion in terms of diagrams converges.
Given a specific instanton coefficient, one can assume that there is some maximum size
for the n-point functions for which the expansion still converges. If the actual
global minimum of the effective action lies beyond the point where the n-point
function reaches this maximum size, the usefulness of this formalism would be
limited.  However, in the case of a continuous phase transition the n-point function
would have a sufficiently small value in the interesting region near the transition.

A semi-classical approximation is made to the action from which the effective action
is derived.  This implies that the theory behaves like a free theory apart from the
presence of a background gauge field.  In addition it is assumed that the instanton
ensemble is fairly dilute -- the average size of the instantons, $\overline{\rho}$,
is smaller than the average separation between them, $R$. On the basis of this
diluteness we drop terms in our derivation that would be suppressed by factors of the
average size over the average separation $(\overline{\rho}/R)$.  As a result our
derivation does not contain interference between zero and non-zeromodes; the overall
non-zeromode determinant factorizes into non-zeromode determinants for each instanton;
and the classical instanton gauge interactions fall away. The instanton vertex for the
fermion zeromodes captures the important dynamics, therefore these
$(\overline{\rho}/R)$ effects are of secondary importance.

The derivation is presented in the following steps.  We start in Section~\ref{nota} by
introducing the necessary source terms in the action, providing some definitions of
the various functionals and defining our notation. The first part of the derivation,
Section~\ref{fermions}, explains how the  fermion fields are treated. An important
part of this involves the treatment of  the zeromodes. The result of this part is a
fermion modal propagator and a  nonlocal fermion vertex for the zeromodes. The next
part, concerning the gauge  fields, is presented in Section~\ref{gauge}. Once all the
quantum fields are  integrated out one is left with an expression that can be
interpreted in terms  of Feynman diagrams.  The relevant Feynman rules are discussed
in Section~\ref{diag}.  The derivation of the effective actions then follows in
Section~\ref{effac}.  It utilizes the procedures of De Dominicis and Martin
\cite{r_dm}. In Section~\ref{2point} we derive the effective action for 2-point
functions and compare it to the CJT effective action \cite{r_cjt}.  Then in
Section~\ref{4point}, starting with the 2-point effective action and setting the mass
part of the full propagator equal to zero, we discuss the derivation of the effective
action for the 4-point functions.  This complicated derivation is in essence identical
to the one that is provided in \cite{r_rth} and its details are therefore not repeated
here.  We conclude with a few important remarks in Section~\ref{concl}.

\section{Notation}
\label{nota}

We are considering the instanton-induced formation of fermion 2-point and
4-point functions in an $SU(N_c)$ gauge theory with $N_f$ massless (Dirac)
fermions in the fundamental representation.  In the semi-classical approximation
we expand the action around the classical background field
configurations, $A_{\mu a}^{(cl)}$:
\begin{equation}
A_{\mu a} = A_{\mu a}^{(cl)} + A_{\mu a}^{(q)},
\label{ie1}
\end{equation}
where $A_{\mu a}^{(q)}$ denotes the ``quantum'' gauge field.  The gauge coupling is
incorporated into the gauge field and appears explicitly in the gauge kinetic term. 
All terms in the action with more than two quantum fields (gauge, ghost or fermion
fields) are dropped.  In the resulting semi-classical action ${\cal S}_2$, the quantum
fields only interact with the background configuration but not with each other.

For the purpose of investigating the formation of 2-point and 4-point functions,
we introduce the following nonlocal 2-fermion source term
\begin{equation}
{\cal S}_K = \int \overline{\psi}(x_1) K(x_1,x_2) \psi(x_2)\ d^4x_1 d^4x_2,
\label{ie2}
\end{equation}
and nonlocal 4-fermion source term
\begin{equation}
{\cal S}_J = \int \overline{\psi}(x_1) \psi(x_2) J(x_1 ...  x_4) \overline{\psi}(x_3)
\psi(x_4)\ d^4x_1 ...  d^4x_4 .
\label{ie3}
\end{equation}
The implicit color and flavor indices are contracted on the sources in (\ref{ie2}) and
(\ref{ie3}).  In the diagrammatic formalism developed here, these nonlocal sources are
treated as 2-fermion and 4-fermion vertices respectively.  We also have the usual
local source terms for the fermion fields:
\begin{equation}
{\cal S}_{\eta} = \int \overline{\eta} \psi \ d^4x + \int \overline{\psi} \eta \ d^4x .
\label{ie4}
\end{equation}

The partition functional $Z$ and generating functional $W$ are defined (in
Euclidean space) as
\begin{equation}
Z = \exp \left( -W[K,J] \right) = \left. \int \exp(-{\cal S}^{(E)})\ {\cal D} 
\right|_{\overline{\eta} = \eta = 0} .
\label{ie5}
\end{equation}
Here, ${\cal D}$ is the functional measure over all the fields in the action,
and the Euclidean action is
\begin{equation}
{\cal S}^{(E)} = {\cal S}_2 + {\cal S}_{\eta} - {\cal S}_K + {\cal S}_J.
\label{ie6}
\end{equation}

The 2-point function is formally given by
\begin{equation}
{\delta W[K,J] \over \delta K} = S[K,J] = \left \langle T \left( \psi
\overline{\psi} \right) \right \rangle
\label{ie7}
\end{equation}
while the 4-point function is
\begin{equation}
{\delta W[K,J] \over \delta J} = G[K,J] = \left \langle T \left( \psi
\overline{\psi}
\psi \overline{\psi} \right) \right \rangle.
\label{ie8}
\end{equation}

We perform the Legendre transformation in two steps.  The first step gives an
effective action for 2-point functions, $\Gamma_2[S,J]$, in the presence of the
nonlocal source, $J$.  The second gives the complete effective action for 2-point and
4-point functions, $\Gamma[S,G]$.  We are however only interested in the case where
fermions remain massless.  Therefore we fix the full propagator to be massless and
denote the resulting effective action for 4-point functions by $\Gamma_4[C]$ in terms
of the {\it amputated} 4-point functions, $C$.

In order to simplify our expressions we replace the fermion fields in the nonlocal
source terms by functional derivatives with respect to the local sources, using the
shorthand
\begin{equation}
\partial = {\delta \over \delta \eta} ~~~ {\rm and} ~~~ \overline{\partial} =
{-\delta \over \delta \overline{\eta}} .
\label{ie9}
\end{equation}
Then these terms are pulled out of the functional integral:
\begin{equation}
Z = \left. \exp \left[ \partial K \overline{\partial} - (\partial
\overline{\partial}) J (\partial \overline{\partial}) \right] Z_0
\right|_{\overline{\eta} = \eta = 0} .
\label{ie10}
\end{equation}
The derivation is presented using the partition functional without the nonlocal
sources,
\begin{equation}
Z_0 = \int \exp \left( - {\cal S}_2 - {\cal S}_{\eta} \right)\ {\cal D} ,
\label{ie11}
\end{equation}
up to the point where the nonlocal sources start to play a role. 

\section{Fermion zeromodes}
\label{fermions}

We first consider the fermion fields.  One can split the action into a fermion part
and a pure gauge part that contains no fermion fields:
\begin{equation}
{\cal S}_2 = {\cal S}_g + \int \overline{\psi} D\!\!\!\!/\ \psi\ d^4x .
\label{ie12}
\end{equation}
The fermion terms in the Lagrangian in $Z_0$ are
\begin{equation}
{\cal L}_{\psi} = \overline{\psi}D\!\!\!\!/\ \psi + \overline{\eta} \psi +
\overline{\psi} \eta ,
\label{ie13}
\end{equation}
where the Dirac operator,
\begin{equation}
D\!\!\!\!/\ = \gamma_{\mu} \left( \partial_{\mu} - i A^{(cl)}_{\mu a} T^a 
\right) ,
\label{ie14}
\end{equation}
contains only the background gauge field, $A^{(cl)}_{\mu a}$.
 
In the presence of instantons (i.e.\ when the background gauge field has a nontrivial
winding number) the Dirac operator possesses fermion zeromodes.  It is singular and
cannot be inverted.  One can avoid this problem by performing a projection onto the
subspace of non-zeromodes and inverting the operator on this subspace only.  For one
instanton with $N_f$ massless fermions there are $N_f$ zeromodes.  Considering a
specific configuration (size, position and color orientation) of this instanton, one
can separate the subspace spanned by the zeromodes from the rest which makes up the
non-zeromodes.  The Dirac operator annihilates the zeromodes, therefore we have
\begin{eqnarray}
{\cal L}_{\psi} & = & \overline{\psi} \left[P_N+P_0\right] D\!\!\!\!/\
\left[P_N+P_0\right] \psi +\overline{\eta} \left[P_N+P_0\right] \psi +
\overline{\psi} \left[P_N+P_0\right] \eta \nonumber \\ & = &
\overline{\psi}_N D\!\!\!\!/\ \psi_N + \overline{\eta}_N \psi_N +
\overline{\psi}_N \eta_N + \overline{\eta}_0 \psi_0 + \overline{\psi}_0 \eta_0 ,
\label{ie15}
\end{eqnarray}
where the operator $P_N$ projects onto the non-zeromodes, while $P_0=1-P_N$ is
the projection operator onto the zeromodes; furthermore $\psi_0 = P_0 \psi $, 
$\psi_N = P_N \psi$, $\eta_0 = P_0 \eta $, $\eta_N = P_N \eta$, etc.

For several instantons and anti-instantons the would-be zeromodes of each one are not
necessarily zeromodes of the total configuration.  In fact the would-be zeromodes mix
and give a spectrum of eigenvalues around zero.  In our approach we keep all the
would-be zeromodes separate from the rest of the fermion modes.  In what follows we
shall proceed to call them `zeromodes' even though  they are not always actual
zeromodes of the total configuration.

Now we define the zeromode and non-zeromode subspaces for any specific background
configuration with multiple instantons. The zeromodes of all the instantons and
anti-instantons span a subspace of the space of all fermion fields.  We separate this
subspace and refer to it as the space of all zeromodes.  The remaining part is called
the space of all non-zeromodes.  Note that the splitting of the space of fermion
fields into zeromodes and non-zeromodes depends on the background configuration.

For ensembles of instantons and anti-instantons the Dirac operator gives a small
but nonzero overlap between the zeromodes of neighboring instantons and
anti-instantons.  Therefore, for instanton ensembles we have, in addition to the terms
in (\ref{ie15}), the terms
\begin{equation}
\overline{\psi}_0^{(I)} D\!\!\!\!/\ \psi_0^{(A)} + \overline{\psi}_0^{(A)}
D\!\!\!\!/\ \psi_0^{(I)},
\label{ie16}
\end{equation}
where the superscripts $(I)$ and $(A)$ indicate that the zeromode belongs
respectively to an instanton or an anti-instanton.

The cross terms of (\ref{ie16}) are responsible for the dynamics that produce the
delocalization of the fermion wave function when chiral symmetry breaking occurs.  It
thus plays a crucial role in the nonperturbative formation of fermion n-point
functions through instantons.  Without the cross terms there would be no fermion
interconnection between instantons and anti-instantons other than through the
interference between zeromodes and non-zeromodes via the Dirac operator.

These interference terms,
\begin{equation}
\overline{\psi}_0 D\!\!\!\!/\ \psi_N + \overline{\psi}_N D\!\!\!\!/\ \psi_0 ,
\label{ie17}
\end{equation}
give contributions to the fermion propagator that mix zeromodes and non-zeromodes. 
However, to leading order in $\overline{\rho}/R$ ($R$ is the average instanton
separation and $\overline{\rho}$~is the average instanton size) the non-singular part
of the fermion propagator does not have terms that connect zeromodes to non-zeromodes
\cite{r_bc}.  This is related to the fact that, to leading order in
$\overline{\rho}/R$, the complete fermion determinant factorizes into a zeromode part
and a non-zeromode part \cite{r_bc,r_lb}.

Our aim is now to find a way to invert the Dirac operator, and arrive at propagators
that correspond to the various types of fermion modes, while avoiding the
singularities.  The remainder of this section addresses this issue.

According to the preceding discussion one can express the partition functional of
(\ref{ie11}) as
\begin{eqnarray}
Z_0 & = & \int {\cal D}_g  \exp \left( - {\cal S}_g - \int \partial_0 D\!\!\!\!/\
\overline{\partial}_0\ d^4x \right)  \nonumber \\ & & \times  \int {\cal D}_{\psi}
\exp\left\{ - \int\left( \overline{\psi}_N D\!\!\!\!/\ \psi_N +
\overline{\eta}_N \psi_N + \overline{\psi}_N \eta_N + \overline{\eta}_0 \psi_0 +
\overline{\psi}_0 \eta_0\ \right)d^4x \right\}.
\label{ie18}
\end{eqnarray}
Here, ${\cal D}_{\psi}$ is the functional measure over the fermion fields and ${\cal
D}_g$ is the functional measure over the gauge fields (quantum plus background) and
the ghost fields. Using $\overline{\partial}_0 = -\delta/ \delta \overline{\eta}_0$
and $\partial_0 = \delta/ \delta \eta_0$, we pulled the cross terms of (\ref{ie16})
out of the functional integral over fermion fields and represented them as one cross
term. It must remain under the gauge functional integral because it still depends on
the background gauge fields through the Dirac operator in (\ref{ie14}).

Integrating over the non-zeromodes, we find
\begin{eqnarray}
Z_0 & = & \int {\cal D}_g \exp \left( - {\cal S}_g - \int \partial_0 D\!\!\!\!/\
\overline{\partial}_0\ d^4x \right) {\rm det}_N\!\left( D\!\!\!\!/\ \right)\
\exp \left( \overline{\eta}_N S_N \eta_N \right) \nonumber \\ & & \times \int {\cal
D}_{\psi_0} \exp \left( - \int\left( \overline{\eta}_0 \psi_0 + \overline{\psi}_0
\eta_0\right)\ d^4x \right) ,
\label{ie19}
\end{eqnarray}
where ${\cal D}_{\psi_0}$ is the measure over the fermion zeromodes and ${\rm
det}_N (D\!\!\!\!/\ )$ is the determinant of the part of the Dirac operator
associated with non-zeromodes, calculated on the subspace of non-zeromodes.  The
non-zeromode propagator $S_N$ is defined by
\begin{equation}
S_N D\!\!\!\!/\ = D\!\!\!\!/\ S_N = P_N ~~~ {\rm and} ~~~ S_N P_0 = P_0 S_N = 0.
\label{ie20}
\end{equation}

The next step is to perform the integration over the fermion zeromodes.
Consider first the case of only one instanton.  At the end we shall generalize
the result to multiple instantons.  We write the two zeromode terms in the
exponent in (\ref{ie19}) as a sum over flavors,
\begin{equation}
\psi_0(x) = \sum_{i=1}^{N_f} \xi^0_i \psi^0_i(x),
\label{ie21}
\end{equation}
where $\xi^0_i$ are Grassmann variables for the zeromodes and $\psi^0_i$ denote
c-number spacetime functions of the fermion zeromodes.  For an explicit form of these
functions see \cite{r_cc}.

Performing the functional integral over zeromodes, we get
\begin{equation}
\int d \!\left[ \overline{\xi}^0, \xi^0 \right] \exp \left\{ - \sum_{i=1}^{N_f}
\int\left( \overline{\eta}_0 \xi^0 \psi^0 + \overline{\psi}^0 \overline{\xi}^0
\eta_0 \right)_i \right\} = \prod_{i=1}^{N_f} \left( \int \overline{\eta}_0\,
\psi^0 \right)_i \left( \int \overline{\psi}^0\, \eta_0 \right)_i .
\label{ie22}
\end{equation}
where $d [\overline{\xi}^0,\xi^0]$ denotes the functional measure over the Grassmann
variables of the fermion zeromodes.  The subscripts $i$ on the brackets indicate that
the enclosed sources, Grassmann variables and zeromode functions are all associated
with the same flavor.  The result is an object that, from a diagrammatic point of
view, behaves like an n-point propagator ($n=2 N_f$). 

It would be more convenient to turn this n-point function into an n-fermion vertex. 
We achieve this by adding the following source terms to the Euclidean action in
(\ref{ie6}):
\begin{equation}
{\cal S}_{\eta_1} = \overline{\eta}_1 S_0 \eta_0 + \overline{\eta}_0 S_0 \eta_1 .
\label{ie23}
\end{equation}
Here each zeromode source $\eta_0$ has associated with it a new source,
$\overline{\eta}_1$, and similarly an $\eta_1$ is associated to $\overline{\eta}_0$.
(Integrations over the relevant spacetime coordinates are understood.)

The addition of these terms is strictly formal.  The two types of sources serve merely
as variables with respect to which the functional derivatives are taken.  In essence,
both types of sources are associated with zeromodes and are in that respect not
different.  However, the way they appear in the expression implies that their dynamics
are different.  The introduction of these source terms changes the partition
functional.  So now we define a new partition functional $Z_1$ which includes these
sources. Upon setting $\overline{\eta}_1 = \eta_1 = 0$ we recover $Z_0$.

The idea is now to amputate the n-point propagator by writing
\begin{eqnarray}
\prod_i \left( \int \overline{\eta}_0 \psi^0 \right)_i \left( \int \overline{\psi}^0
\eta_0 \right)_i \exp \left( {\cal S}_{\eta_1} \right) = \prod_i \left( \partial_1
S_0^{-1} \psi^0 \right)_i \left( \overline{\psi}^0 S_0^{-1} \overline{\partial}_1
\right)_i \exp \left( {\cal S}_{\eta_1} \right).
\label{ie24}
\end{eqnarray}
Here we use the functional derivatives $\overline{\partial}_1 = -\delta/ \delta
\overline{\eta}_1$ and $\partial_1 = \delta/ \delta \eta_1$.  The subscripts $i$ on
the right-hand side indicate that both the functional derivatives and the zeromode
functions inside the brackets are associated with the same flavor.

So now we have
\begin{eqnarray}
Z_1 & = & \int {\cal D}_g \left[ \exp \left( - {\cal S}_g - \int \partial_0
D\!\!\!\!/\ \overline{\partial}_0\ d^4x \right) {\rm det}_N\!\left( D\!\!\!\!/\
\right) \right. \nonumber \\ & & \times\ \left. \prod_i^{N_f} \left( \partial_1
S_0^{-1} \psi^0 \right)_i \left( \overline{\psi}^0 S_0^{-1} \overline{\partial}_1
\right)_i \exp \left( \overline{\eta}_1 S_0 \eta_0 + \overline{\eta}_0 S_0
\eta_1 + \overline{\eta}_N S_N \eta_N \right) \right].
\label{ie25}
\end{eqnarray}
The generalization of (\ref{ie25}) for multiple instantons and anti-instantons is
straightforward.  The only difference is that the number of zeromodes increases. When
the background configuration contains $n$ instantons and $m$ anti-instantons the
product of all zeromodes in (\ref{ie25}) runs over
$(n+m)N_f$ instead of $N_f$ factors.

Now we allow the cross term,
\begin{equation}
\exp \left( - \int \partial_0 D\!\!\!\!/\ \overline{\partial}_0\ d^4x \right) ,
\label{ie26}
\end{equation}
to operate on the propagators
\begin{equation}
\exp \left( \overline{\eta}_1 S_0 \eta_0 + \overline{\eta}_0 S_0 \eta_1 +
\overline{\eta}_N S_N \eta_N \right).
\label{ie27}
\end{equation}
The vertex, $\partial_0 D\!\!\!\!/\ \overline{\partial}_0$, connects zeromodes of
instantons to anti-instantons while the bare propagator, $S_0$, is associated with
zeromodes of a specific instanton or anti-instanton. One can use the vertex to get a
propagator that connects the zeromodes of different instantons:
\begin{equation}
\exp \left( - \int \partial_0 D\!\!\!\!/\ \overline{\partial}_0\ d^4x \right) \exp
\left( \overline{\eta}_1 S_0 \eta_0 + \overline{\eta}_0 S_0 \eta_1 \right) = \exp
\left( \overline{\eta}_1 S_Z \eta_1 + \overline{\eta}_1 S_0 \eta_0 + \overline{\eta}_0
S_0 \eta_1 \right) .
\label{ie28}
\end{equation}
Here $S_Z = - S_0 (i D\!\!\!\!/) S_0$ propagates zeromodes between instantons and
anti-instantons (the negative sign is due to the Grassmann nature of the sources). 
The bare propagator $S_0$ connects zeromodes between instantons and source vertices.

In order to simplify the notation we combine all the propagators into one,
\begin{equation}
\exp \left( \overline{\eta}_1 S_Z \eta_1 + \overline{\eta}_1 S_0 \eta_0 +
\overline{\eta}_0 S_0 \eta_1 + \overline{\eta}_N S_N \eta_N \right) = \exp \left(
\overline{\eta}^{\prime} S_m \eta^{\prime} \right) ,
\label{ie29}
\end{equation}
and call it ($S_m$) the modal propagator.  Its various parts, as shown in (\ref{ie29}),
depend on the background configuration in a complicated way. However, the expressions
for these parts are known \cite{r_bc,r_lb,r_bccl}.

The resulting expression of the partition functional is
\begin{equation}
Z_1 = \int {\cal D}_g \left[ \exp \left( - {\cal S}_g \right) {\rm det}_N \left(
D\!\!\!\!/\ \right) \prod_i^{n_z} \left( \partial_1 S_0^{-1} \psi^0 \right)_i \left(
\overline{\psi}^0 S_0^{-1} \overline{\partial}_1 \right)_i \exp \left(
\overline{\eta}^{\prime} S_m \eta^{\prime} \right) \right] ,
\label{ie30}
\end{equation}
where $n_z$ denotes the total number of zeromodes.

We have now completed the derivation of the fermion propagator. All fermion fields
have been successfully integrated out and we have a single propagator for the fermion
fields. In addition we have a nonlocal $2 N_f$-fermion vertex that describes the
interactions of fermion zeromodes.

\section{Gauge fields}
\label{gauge}

At this point the only remaining quantum fields are the ghost and gauge boson fields
that appear in the gauge action ${\cal S}_g$.  In the semi-classical approximation the
functional measure ${\cal D}_g$ can be split into a functional measure over the
remaining quantum fields, ${\cal D}_q$, and the measure over the collective
coordinates (CC's) that specify the background field configuration. We denote the
collective coordinates by $\zeta$.

To evaluate the functional integral over the quantum fields, we split the gauge action
into two parts:\ ${\cal S}_g={\cal S}_t+{\cal S}_{int}$.  Here ${\cal S}_t$ consists
of the kinetic terms for the ghosts and gauge bosons and the sum of the classical
actions of the individual instantons.  The functional integral of the part containing
${\cal S}_t$ leads to the 't Hooft amplitude without fermions \cite{r_t}.  The part of
the gauge action, ${\cal S}_g$, that contains mixtures of different instantons and
anti-instantons behaves like an interaction terms and is therefore denoted by ${\cal
S}_{int}$.  The latter contains no quantum fields.  After integrating over the color
orientations of the instantons the surviving part of ${\cal S}_{int}$ is of
$O([\overline{\rho}/R]^6)$ \cite{r_dp}.  We therefore drop ${\cal S}_{int}$ from
our analysis.

The fermion contribution in the 't Hooft amplitude can be separated into a zeromode
and a non-zeromode part.  The non-zeromode part comes from the non-zeromode
determinant which first appeared in (\ref{ie19}).  For $N_f$ fermions with mass $m$,
the zeromodes give a factor of $(m/\mu)^{N_f}$, where $\mu$ is the renormalization
scale.  In the case of massless fermions, one needs additional fermion dynamics in the
theory to take care of the zeromodes.  Without this additional dynamics the instanton
amplitude for massless fermions would vanish.  In our case this dynamics is provided
by the nonlocal sources or, in the effective action language, by the n-point functions.

To leading order in $\overline{\rho}/R$ the non-zeromode part of the fermion
determinant factorizes into separate non-zeromode determinants for each instanton
\cite{r_bc,r_lb}. For $n$ instantons and $m$ anti-instantons we have:
\begin{equation}
{\rm det}_N \left( D\!\!\!\!/\ \right) = \prod_i^{n+m} {\rm det}_N \left(
D\!\!\!\!/_i\ \right) + O \left({\overline{\rho}\over R}\right) ,
\label{ie31}
\end{equation}
where $D\!\!\!\!/_i$ denotes a Dirac operator with only one instanton in the
background field configuration.

We shall first discuss the integration over quantum gauge fields for one instanton and
then at the end generalize the result for multiple instantons. Apart from the fermion
dynamics that must account for the fermion zeromodes, the 't Hooft vacuum-to-vacuum
tunneling amplitude for one instanton is given by
\begin{equation}
{\cal A} = {1\over\rm \cal N}\, {\rm det}_N \left( D\!\!\!\!/\ \right) \int \exp
\left( - {\cal S}_t \right)\ {\cal D}_q,
\label{ie32}
\end{equation}
where ${\cal N}$ is the product of the same determinant and gauge functional integral
evaluated in the absence of any instantons.  Hence ${\cal N}$ is a collective coordinate independent
normalization constant.

The vacuum-to-vacuum tunneling amplitude in (\ref{ie32}) becomes \cite{r_t}
\begin{equation}
{\cal A} = \kappa \gamma(\alpha) \frac{1}{\rho^5} (\mu\rho)^{b_0+N_f},
\label{ie33}
\end{equation}
where the dependence on the gauge coupling is contained in
\begin{equation}
\gamma \left( \alpha(\mu) \right) = \left( {2 \pi \over \alpha(\mu)} \right)^{2
N_c} \exp \left(- {2 \pi \over \alpha(\mu)} \right) .
\label{ie34}
\end{equation}
Here $\mu$ is the renormalization scale and $b_0 = \frac{1}{3}(11 N_c - 2 N_f)$.  The
running of the coupling in (\ref{ie33}) is determined by $\gamma(\alpha)
(\mu\rho)^{b_0}$.  As such it gives one-loop running for the coupling that appears in
the exponent in $\gamma(\alpha)$.  The expression can be extended to higher orders
which would cause the coupling in the monomial-part of $\gamma(\alpha)$ to run as
well. The prefactor in (\ref{ie33}) is
\begin{equation}
\kappa = {2 \exp(5/6) \exp(B N_f-C N_c]) \over \pi^2 (N_c-1)! (N_c-2)!}
\label{ie35}
\end{equation}
with $B=0.2917$, $C=1.5114$ to one loop in MS-bar \cite{r_l}.  Here $\kappa$ includes a
color volume factor that comes from the integral over all color orientations.  This
integral only gives the volume of the relevant factor group, $SU(N_c)/G$, where $G$ is
the subgroup that leaves the instantons invariant.  This prefactor does not include
the effect of the gauge dependence in the remaining dynamics.  The complete integral
over color orientations can be seen as a product of a color volume factor and an
averaging over the gauge structure of the remaining dynamics.  The latter, which
we consider next, gives a tensor structure that indicates how color indices
are contracted.

The gauge dependence of the remaining dynamics appears in the color orientation of the
fermion zeromodes.  One can represent this as
\begin{equation}
\prod_{i=1}^{N_f} \left( \partial_1 g S_0^{-1} \psi^0 \right)_i \left(
\overline{\psi}^0 S_0^{-1} g^{\dag} \overline{\partial}_1 \right)_i
\label{ie36}
\end{equation}
where $g$ and $g^{\dag}$ are gauge transformations operating on the zeromode functions
to give all possible color orientations. The integral over all color orientations can
therefore be written as 
\begin{equation}
\int \prod_{i=1}^{N_f} \left( g_{a_i b_i}  g^{\dag}_{c_i d_i} \right) dg
\label{ie37}
\end{equation}
where $dg$ is the Haar measure for the gauge group in which the SU(2) group of
the instanton is embedded.  The result of the integral in (\ref{ie37}) for an
arbitrary number of flavors, $N_f$, is quite complicated.  However, it
simplifies significantly in the case of a large number of colors, $N_c$, when
one neglects the terms that are subleading in $1/N_c$.  The result is
\begin{equation}
\int \prod_{i=1}^{N_f} \left( g_{a_i b_i} g^{\dag}_{c_i d_i} \right) dg \approx
N_c^{-N_f} \left[ \prod_{i=1}^{N_f} \left( \delta_{a_i d_i} \delta_{b_i c_i}
\right) + (c,d){\rm permutations} \right] ,
\label{ie38}
\end{equation}
where the $c_i$ and $d_i$-indices are permutated together.  Permuting the flavor
indices instead of color indices, one finds that if such a permutation is odd the term
receives a negative sign due to the Grassmann nature of the functional derivatives. 
The flavor permutations therefore lead to an antisymmetric
$\epsilon$-tensor:
\begin{equation}
N_c^{-n} \epsilon^{g_1 g_2 ...  g_n} \prod_{i=1}^n \left( \partial_1^{i a_i}
S_0^{-1} \psi^0_{i b_i} \right) \left( \overline{\psi}^0_{g_i b_i} S_0^{-1}
g^{\dag} \overline{\partial}_1^{g_i a_i} \right) .
\label{ie39}
\end{equation}
Here $n$ is the number of flavors.  The first index on each zeromode or derivative is
a flavor index and the second one is a color index.  Note that the flavors of the
first bilinears are just those produced by the index of the product, while those on
the second bilinears are contracted with the $\epsilon$-tensor, and thus give a
permuted version of the flavors on the first bilinears.  One can see that for only two
flavors this expression resembles the first term of the 't Hooft vertex in \cite{r_t}.

We now write (\ref{ie30}) for the one instanton case as
\begin{equation}
Z_1 = \int {\cal V}_I \exp \left( \overline{\eta}^{\prime} S_m \eta^{\prime}
\right)\ d\zeta ,
\label{ie40}
\end{equation}
where the object ${\cal V}_I$ plays the role of a nonlocal vertex:
\begin{equation}
{\cal V}_I = {\cal A} N_c^{-N_f} \epsilon^{\{g\}} \prod_i^{N_f} \left(
\partial_1 S_0^{-1} \psi^0 \right)_i \left( \overline{\psi}^0 S_0^{-1}
\overline{\partial}_1 \right)_i ,
\label{ie41}
\end{equation}
with $\{g\}={g_1 g_2 ...  g_{N_f}}$.  A similar expression exists for
anti-instantons, which we denote by ${\cal V}_A$.  The only difference is that
the zeromode functions, $\psi^0$ and $\overline{\psi}^0$, have the opposite
helicities.

The quantum gauge and ghost fields interact with one instanton at a time.  For this
reason, when we generalize to configurations with many instantons, we get a factor of
${\cal A}$ for each instanton.  There is a ${\cal V}_I$ or ${\cal V}_A$, for each
respective instanton and anti-instanton in the background configuration.  At the
same time each instanton comes with a complete set of collective coordinates.

The resulting expression is
\begin{equation}
Z_1 = \sum_{n=0}^{\infty} \sum_{m=0}^{\infty} {1\over n!} {1\over m!} \left(
\prod^n_{i=1} \int d\zeta_i\ {\cal V}_{Ii} \right) \left( \prod^m_{j=1} \int
d\zeta_j\ {\cal V}_{Aj} \right) \exp \left( \overline{\eta}^{\prime} S_m
\eta^{\prime} \right) ,
\label{ie42}
\end{equation}
where each ${\cal V}_I$ and ${\cal V}_A$ introduces its own integral over its
associated collective coordinates.  The understanding is that $\exp(-i
\overline{\eta}^{\prime} S_m \eta^{\prime})$ is part of the integrand of the
overall collective coordinate integral.

We now recover the original partition functional with the nonlocal sources using
\begin{equation}
Z = \left. \exp \left[ \partial K \overline{\partial} - (\partial
\overline{\partial}) J (\partial \overline{\partial}) \right] Z_1
\right|_{\overline{\eta} = \eta = \overline{\eta}_1 = \eta_1 = 0} .
\label{ie43}
\end{equation}

\section{A diagrammatic language}
\label{diag}

In \cite{r_rth,r_dm,r_cjt} the diagrammatic language of the underlying gauge theory was
exploited to express the effective action as a functional of the relevant n-point
functions.  For instanton physics the necessary diagrammatic language is provided
by the Feynman rules that can be derived from the expressions for $Z$ in (\ref{ie42})
and (\ref{ie43}). These include rules for:
\begin{itemize} 
\item the 2-fermion and 4-fermion sources, $K$ and $J$, that are represented by
nonlocal 2-fermion and 4-fermion vertices in the diagrams;
\item the nonlocal $2 N_f$-point instanton vertices, ${\cal V}_I$ and ${\cal V}_A$;
\item the modal propagator $S_m$, represented by fermion lines, and
\item an integral over the collective coordinates of all the instantons involved in
the diagram.
\end{itemize}

According to these rules $Z[K,J]$ is the sum of all vacuum diagrams in the presence of
the collective coordinate independent nonlocal sources $K$ and $J$.  The modal
propagator depends on all the collective coordinates of the background configuration.
As a result disconneted diagrams do not factorize. However, if one assumes that the
instanton ensemble is fairly dilute, the overall collective coordinate integral can be
separated into a product of integrals -- one for each connected part of the diagram.
This is because the modal propagator depends only weakly on the collective coordinates
of far away instantons.  Under the above assumption each connected diagram only
depends on the instantons that are directly involved, so that $Z$ exponentiates.
Therefore $W = - \ln(Z)$ consists of the sum of all (collective coordinate integrated)
{\it connected} vacuum diagrams.

\section{The effective actions}
\label{effac}

The effective action is the Legendre transform of the sum of all connected
vacuum diagrams with respect to the nonlocal sources.  The Legendre transform
formally replaces the dependence on these sources by a dependence on the n-point
functions.  However, to find the explicit expressions one must perform a
resummation of the vacuum diagrams in terms of the n-point functions.  This
resummation hides the sources inside the n-point functions except in the term
that is explicitly removed by the Legendre transform.  Thus it leads to an
expression for the effective action in terms of these n-point functions.

De Dominicis and Martin \cite{r_dm} provided diagrammatic procedures for the
systematic resummation of vacuum diagrams in terms of the relevant n-point
functions (up to 4-point functions).  We use these procedures to derive the
effective actions for 2-point and 4-point functions.  While in the 2-point case
we allow chirality breaking structures that can generate dynamical masses, in
the 4-point case we specialize to situations in which unbroken chiral symmetries
protect fermions from acquiring masses.

\subsection{The 2-point effective action}
\label{2point}

\begin{figure} 
\centerline{\epsfysize = 10 cm \epsfbox{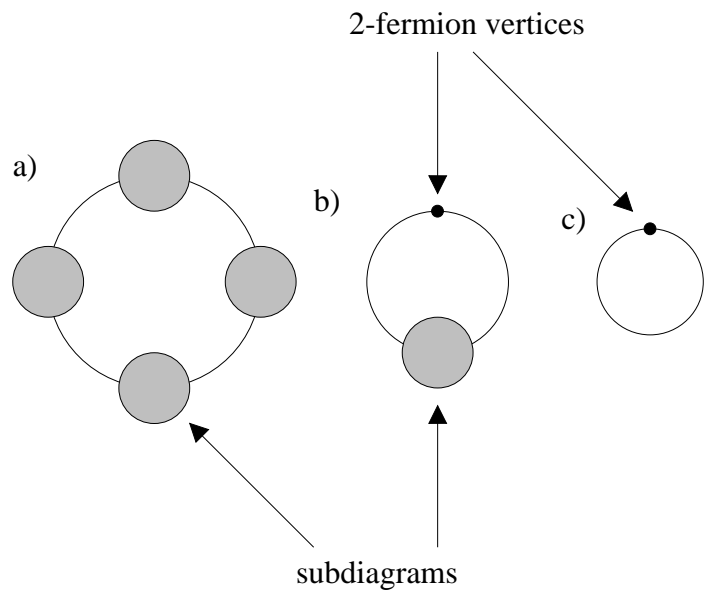}}
\caption{Examples of cycles of lines: a) a 4-cycle, b) a trivial 2-cycle and c) a
trivial 1-cycle.}
\label{lusse}
\end{figure}

We first define a few useful concepts \cite{r_dm}.  A {\it cycle of lines} or an {\it
m-cycle} in a vacuum diagram is a set of $m$ fermion lines with the property that when
we cut all these $m$ lines the vacuum diagram separates into $m$ disjoint 2-point
diagrams (see Figure \ref{lusse}a).  When one and only one of these 2-point diagrams
consists of only one 2-fermion vertex the cycle of lines is called {\it trivial} with
respect to this type of 2-fermion vertex (see Figure \ref{lusse}b and c).

Next we define the (connected, collective coordinate integrated) {\it 2-particle irreducible} (2PI)
vacuum diagrams as those vacuum diagrams that contain only nontrivial 1-cycles and
trivial 2-cycles with respect to the 2-fermion vertex $K$.  (Note that the only
trivial 1-cycle, Tr$\{ S_m\, K\}$, shown in Figure \ref{lusse}c, is thus excluded.) 
The sum of all 2PI vacuum diagrams is denoted by $W_{2PI}$.

One can now readily define the (connected, collective coordinate integrated) 2PI {\it 2-point} diagrams
in terms of the 2PI vacuum diagrams through a functional derivation in the
2-fermion vertex $K$.  The sum of all amputated 2PI 2-point diagrams is given by
\begin{equation}
\Sigma[S_m,J] = \left. {\delta W_{2PI}[S_m,J] \over \delta K} \right|_{amp}.
\label{ie44}
\end{equation}
In this expression we show the dependence on the propagator to indicate that the lines
in these diagrams represent the modal propagator, $S_m$.  We note that $\Sigma[S_m,J]$
has the same collective coordinate integrals that $W_{2PI}[S_m,J]$ has, because $K$ is independent
of the collective coordinates.

One can construct the sum of all 2-point diagrams from the sum of all 2PI 2-point
diagrams, $\Sigma[S_m,J]$, by simply replacing the modal propagator with the full
propagator $S$.  Thus $\Sigma[S,J]$ represents the sum of all 2-point
diagrams.\footnote{The diagrams in $\Sigma[S,J]$ are by definition 2PI with respect to
the full propagator $S$. However when the full propagator inside these diagrams are
replaced by the sum of diagrams in which fermion lines represent $S_m$, the diagrams
in $\Sigma[S,J]$ are not necessarily 2PI anymore.}

Now we introduce the topological equation \cite{r_dm}:
\begin{equation}
1 = N_{cycles} - N_{lines} + N_{skel} ,
\label{ie45}
\end{equation}
where $N_{cycles}$ is the number of cycles of lines in a vacuum diagram, $N_{lines}$
is the total number of lines in the diagram, and $N_{skel}$ is the number of 2PI
skeletons.  A 2PI skeleton is what remains of a 2PI vacuum diagram after one has cut
out all 2-fermion vertices.  The topological equation (\ref{ie45}) holds for every
vacuum diagram separately.

One can use this topological equation to construct a resummation equation for
the sum of all vacuum diagrams. This resummation equation, which will make it 
possible to write the sum of all vacuum diagrams in terms of the 2-point
functions, is given by
\begin{equation}
W = W_{cycles} - W_{lines} + W_{skel}.
\label{ie46}
\end{equation} 
Each term on the right-hand side of this equation is formed by counting each
vacuum diagram as many times as the number of the corresponding elements
(cycles, lines or skeletons) that appear in it. According to (\ref{ie46}) these
three terms give back the sum of all vacuum diagrams, $W$.

One forms $W_{cycles}$ by constructing all cycles:
\begin{equation}
W_{cycles} = {\rm Tr} \left\{ \sum_{n=1}^{\infty} \frac{-1}{n} \left( S_m\,
\Sigma[S,J] \right)^n \right\} = {\rm Tr} \left\{ \ln \left( 1 - S_m\,
\Sigma[S,J] \right) \right\}.
\label{ie47}
\end{equation}
The minus sign inside the summation is because of the fermion loop. To form
$W_{lines}$ one closes off the sum of all 2-point diagrams (minus the modal
propagator) with
$S_m^{-1}$:
\begin{equation}
W_{lines} = - {\rm Tr} \left\{ S_m^{-1} \left( S - S_m \right) \right\} .
\label{ie48}
\end{equation}
The minus sign in front is because we are closing off a fermion loop. The term
$W_{skel}$ is formed by replacing all the $S_m$'s inside the sum of all 2PI vacuum
diagrams, $W_{2PI}$, by the full propagator $S$ and add to this Tr$\{S K\}$, because
$K$ is also a 2PI skeleton and~Tr$\{ S_m\, K\}$~is excluded from the definition of 2PI
vacuum diagrams. So we have
\begin{equation}
W_{skel} = W_{2PI}[S,J] + {\rm Tr} \left\{ S\, K \right\}.
\label{ie49}
\end{equation}

When we place (\ref{ie47}), (\ref{ie48}) and (\ref{ie49}) into (\ref{ie46}) the result
is
\begin{equation}
W = {\rm Tr} \left\{ \ln \left( 1 - S_m\, \Sigma[S,J] \right) \right\} + {\rm Tr}
\left\{ S_m^{-1} \left( S - S_m \right) \right\} + W_{2PI}[S,J] + {\rm Tr} \left\{ S K
\right\}.
\label{ie50}
\end{equation}
One can simplify this by using the following expression that relates $\Sigma$
and $S$,
\begin{equation}
S = S_m + S_m \Sigma[S,J] S ,
\label{ie51}
\end{equation}
which represents the resummation of self-energy diagrams. Then we have 
\begin{equation}
W = {\rm Tr} \left\{ \ln \left( S^{-1} \right) - \ln \left( S_m^{-1} \right) \right\}
+ {\rm Tr} \left\{ \left( S_m^{-1} - S^{-1} \right) S \right\} + W_{2PI}[S,J] + {\rm
Tr} \left\{ S\, K \right\}.
\label{ie52}
\end{equation}
The Legendre transform for 2-point functions removes the last term, leaving us
with an effective action for 2-point functions 
\begin{equation}
\Gamma_2[S,J] = {\rm Tr} \left\{ \ln \left( S^{-1} \right) - \ln \left(
S_m^{-1} \right) \right\} + {\rm Tr} \left\{ \left( S_m^{-1} - S^{-1} \right) S
\right\} + W_{2PI}[S,J],
\label{ie53}
\end{equation}
The 2PI vacuum diagrams in $W_{2PI}[S,J]$ are constructed with Feynman rules that
differ from those mentioned in Section~\ref{diag} only in that there is no 2-fermion
nonlocal source vertex anymore and the modal propagator $S_m$ is now replaced by the
full propagator $S$. The other terms in (\ref{ie53}) are referred to as the `one-loop'
terms.

This expression resembles the CJT effective action \cite{r_cjt},
\begin{equation}
\Gamma_{CJT}[S] = {\rm Tr} \left\{ \ln \left( S^{-1} \right) \right\} - {\rm Tr}
\left\{ \left( S^{-1}-S_0^{-1} \right) S \right\} + W_{2PI}[S] .
\label{ie54}
\end{equation}
The differences are that $\Gamma_2[S,J]$ contains the modal propagator $S_m$, instead
of the bare propagator $S_0$; the traces, Tr$\{\cdot\}$, in $\Gamma_2[S,J]$ include an
integration over the collective coordinates, and $\Gamma_2[S,J]$ includes the term
${\rm Tr} \{\ln(S_m^{-1})\}$, which, after the collective coordinate integration,
becomes a constant that is removed through normalization.

Taking the functional derivative of $\Gamma_2[S,J]$ with respect to the full
propagator, one obtains a gap equation for the full propagator:
\begin{equation}
{\delta \Gamma_2[S,J] \over \delta S} = 0 = S^{-1} - \int S_m^{-1} d\zeta +
{\delta W_{2PI}[S,J] \over \delta S} .
\label{ie55}
\end{equation}
The collective coordinate integral of $S_m^{-1}$ in momentum space equals the bare
propagator multiplied by a momentum-dependent scalar form factor. If one restricts
$W_{2PI}$ to be the one--instanton (and one--anti-instanton) amplitude and considers
the small mass limit, one can show that the gap equation in (\ref{ie55}) reproduces
the Carlitz and Creamer gap equation \cite{r_cc} to leading order in small mass.

\subsection{The 4-point effective action}
\label{4point}

The generation of 4-point functions by instantons is elucidated by the 
complete effective action which is found by performing a second Legendre 
transform:
\begin{equation}
\Gamma[S,G] = \Gamma_2[S,J] - \int J G .
\label{ie56}
\end{equation}
This operation actually only affects the sum of all 2PI vacuum diagrams,
$W_{2PI}[S,J]$, in (\ref{ie53}).  The remaining terms in that expression are
independent of $J$.

Here, as explained in Section~\ref{intro}, we are interested in the case when the
fermions remain massless.  By fixing the full fermion propagator to one that has no
mass term, one can distinguish right-handed and left-handed fermion fields.  The
nonlocal 4-fermion source term in (\ref{ie3}) is for this purpose replaced by
 \begin{eqnarray}
S_J & = & \int \left( \frac{1}{4} \overline{\psi}_L \psi_R J_C \overline{\psi}_L
\psi_R + \frac{1}{4} \overline{\psi}_R \psi_L J_C^{\dag} \overline{\psi}_R
\psi_L + \overline{\psi}_R \psi_L J_0 \overline{\psi}_L \psi_R \right.
\nonumber \\ & & \left.  + \frac{1}{4} \overline{\psi}_R \psi_R J_R
\overline{\psi}_R \psi_R + \frac{1}{4} \overline{\psi}_L \psi_L J_L
\overline{\psi}_L \psi_L \right)\ d^4x_1 ...  d^4x_4 .
\label{ie57}
\end{eqnarray}
These five chirality structures are the only ones that can be included without
completely breaking the chiral symmetry and thereby generating a dynamical mass for
the fermions.  In this expression color, flavor and Dirac indices are implicitly
contracted on the sources.  It turns out to be more convenient to work with the five
{\it amputated} 4-point functions that have the same chiral structures as the sources
in (\ref{ie57}).  The chirality changing 4-point functions, $C_C$ and $C_C^\dagger$ and
the chirality conserving 4-point functions, $C_0$, $C_L$ and $C_R$ are used instead of
the corresponding $G$'s.  The resulting 4-point effective action, $\Gamma_4[C]$, is
stationary with respect to variations in the $C$'s.  These stationarity conditions
coincides with those of $\Gamma[S,G]$ in $G$, because $S$ is kept fixed.

In \cite{r_rth} the procedures of \cite{r_dm} are adapted to gauge theories and
a 4-point effective action is obtained in terms of the five amputated $C$'s:
\begin{eqnarray}
\Gamma_4 & = & {\rm Tr}\left \{ \ln \left(1 - \frac{1}{4} R_s C_C^{\dag} L_s C_C
\right)_s + \ln \left(1 +\frac{1}{2} C_R \right)_s + \ln \left(1 + \frac{1}{2} C_L
\right)_s + \ln(1+C_0)_s \right. \nonumber \\ & & \hspace{3em} + \frac{1}{2} \ln
\left(1 - R C_0 L C_0\right)_d+\frac{1}{2}\ln\left(1+C_R\right)_d + \frac{1}{2} \ln
\left(1 + C_L \right)_d + \ln \left(1 + C_0 \right)_d \nonumber \\ & & \hspace{3em} +
\frac{1}{2} \ln\left(1 - Z C_C^{\dag} Z C_C \right)_d - C_R - C_L - 2 C_0 \left. +
\frac{1}{2} C_C C_C^{\dag} + C_0^2 + \frac{1}{4} C_R^2 + \frac{1}{4} C_L^2 \right \}
\nonumber\\ &&+ W_{4PI}[C_C,C_C^{\dag},C_0,C_R,C_L] .
\label{ie58}
\end{eqnarray}
The derivation of (\ref{ie58}) is based on diagrammatic arguments from
\cite{r_rth,r_dm} which are also applicable to the instanton case at hand.  The only
difference in this case is the set of Feynman rules, which are discussed below. We do
not reproduce the derivation of (\ref{ie58}) here. The one-loop terms of (\ref{ie53})
are dropped from the 4-point effective action because they do not contain $C$'s and
would therefore not affect the 4-point gap equations.

The first term in (\ref{ie58}) contains the result of the summation of 4-particle {\it
reducible} diagrams, i.e.\ those that can be separated into two disjoint parts by
cutting two pairs of fermion lines. The two fermion lines in these pairs can flow in
the same direction ($s$-type) or in opposite directions ($d$-type).  The products of
$C$'s and other operators in (\ref{ie58}) indicate that the associated 4-point
functions are connected to each other by a pair of fermion lines of either the
$s$-type or $d$-type.  The type of fermion pair involved is indicated by the subscript
on the brackets.  The other operators in (\ref{ie58}) are defined as follows in terms
of sequences of $C$'s:
\begin{eqnarray} 
& Z = \left( 1+C_0 \right)_d^{-1}, ~~ L = \left( 1+C_L \right )_d^{-1}, ~~ R =
\left( 1+C_R \right)_d^{-1} & \nonumber \\ & L_s = \left( 1+\frac{1}{2}C_L
\right)_s^{-1}, ~~ R_s = \left( 1+\frac{1}{2}C_R \right)_s^{-1} . &
\label{ie59}
\end{eqnarray}

The last term in (\ref{ie58}), $W_{4PI}$, is the sum of all 4PI\footnote{The definition of 4PI is such that the
possibility to remove {\it only} a $C$ by cutting four lines in a diagram does not
imply that the diagram is not 4PI.  However, the diagram consisting of two
interconnected $C's$ is not 4PI.  See \cite{r_rth,r_dm} for the exact definition of
4PI.} vacuum diagrams,
calculated with the relevant Feynman rules. These Feynman rules differ from those
mentioned in Section~\ref{diag} in that there are no nonlocal source vertices anymore.
Instead the 4-point functions now act as 4-fermion vertices. Furthermore, the modal
propagator is replaced by the full propagator. We still have the $2 N_f$-point
instantons vertices, ${\cal V}_I$ and ${\cal V}_A$ and each diagram must still be
integrated over the collective coordinates of all instantons in the diagram. 

The term $W_{4PI}$ contains an infinite set of diagrams.  Therefore any specific
calculation would require a consistent truncation scheme, similar to those employed
in applications of the Dyson-Schwinger equation.

\section{Conclusions}
\label{concl}

The instanton effective action formalism presented here makes it possible to
investigate the nonperturbative formation of 2-point and 4-point functions.  In this
way one can determine whether and under what circumstances these n-point functions can
act as order parameters that signal the (partial) breaking of a chiral symmetry.  Our
effective actions are derived using the diagrammatic procedures developed in
\cite{r_rth,r_dm}.  These diagrammatic procedures require as input a diagrammatic
language, Feynman rules, in terms of which the diagrams can be expressed. We have
presented here such a diagrammatic language for instantons.  The
key features of it are:\
\begin{itemize}
\item the nonlocal $2 N_f$-point instanton vertices, which can be seen as the natural
generalization of the effective 't Hooft vertex to high energy scales;
\item the modal propagator that propagates the different fermion modes, including the
zeromodes, while avoiding the singularities of the Dirac operator.
\end{itemize}

One success of this formalism that is already apparent without doing any specific
calculations is the fact that the 2-point effective action  reproduces the Carlitz and
Creamer gap equation \cite{r_cc} in the small mass limit.

It is perhaps useful to compare this formalism to the instanton liquid model
\cite{r_d,r_s}.  The most obvious difference is the fact that the instanton liquid
formalism is a statistical physics approach whereas the formalism here is a quantum
field theory approach.  The instanton liquid model contains statistical quantities
like average instanton size and average instanton separation.  The effective action
formalism, on the other hand, is formulated in terms of the resummation of Feynman
diagrams containing the instanton dynamics in terms of vertices and propagators.

Another obvious difference is the fact that the instanton liquid approach specifically
includes the instanton interactions, ${\cal S}_{int}$, which we dropped.  Due to this
it might appear that our formalism neglects an important part of the physics that
governs the behavior of instantons.  The point is that the instanton liquid model
addresses a different regime than what we intend to address through our formalism. 
Instanton liquid models are mostly employed in the low energy phenomenology of QCD. 
In this regime the coupling constant is fairly large and the instanton ensemble,
although dilute enough to use a semi-classical approximation, is not dilute enough to
neglect instanton interactions.  For example, Dyakonov and Petrov pointed out in
\cite{r_dp} that the coefficient of the instanton interaction term is large in the
absence of fermions.  One of the crucial aspects in their analysis is the
stabilization of the instanton ensemble.  The instanton interactions, which are on
average repulsive, provide the mechanism for this stabilization.

In contrast to this, the effective action formalism attempts to answer questions about
the generation of order parameters.  These are expected to appear at high energy
scales when the coupling constant is still fairly small.  Furthermore, if the phase
transitions associated with these symmetry breakings are continuous -- second order
or higher -- the order parameters would be small close to the transition point.  The
order parameters are therefore ideal expansion parameters.  Terms containing small
powers of the order parameters will dominate the analysis. The diluteness factor
$(\overline{\rho}/R)$ is related to the tunneling amplitude, which contains positive
powers of the order parameter. One can thus see that terms with factors of diluteness
come with high powers of the order parameter and are therefore suppressed in the region
where these order parameters are small.

Yet it must be acknowledged that the diluteness terms do represent important physics.
This can be seen from the fact that without them the effective potential would not be
bounded below.\footnote{It is a well known fact\cite{r_br} that the 2-point effective
potential -- which in our case does not differ significantly from the CJT effective
potential -- is unbounded below even without dynamics.  This is a problem related to
the nonlocality of the 2-fermion sources.} The dynamics given by the instanton vertices
is purely attractive and by increasing the size of the order parameter one lowers the
energy of the potential. The repulsiveness of the diluteness terms are required to
make the potential bounded from below. This aspect is however irrelevant for the
investigation into the nature of the second order phase transitions that interests us.

It is in principle possible to include the diluteness terms in the effective action
formalism.  One simply needs to derive ``Feynman rules'' for these types of
interactions.  However, the resulting formalism would be awkward. One of the reasons
for neglecting them is because they depend on statistical physics quantities.  Their
inclusion would therefore lead to a mixed statistical physics--quantum field theory
formalism.

\section*{Acknowledgements}

We are grateful to Bob Holdom for helpful discussions. One of us (FSR) also wants
to thank Stephen Selipsky for explaining certain aspects of instantons to him.

\newpage

\end{document}